\documentclass[american,english]{IEEEtran}
\usepackage[T1]{fontenc}
\usepackage[latin9]{inputenc}
\usepackage{xcolor}
\usepackage{amstext}
\usepackage{graphicx}
\usepackage{setspace}
\usepackage{esint}
\PassOptionsToPackage{normalem}{ulem}
\usepackage{ulem}
\makeatletter

\providecommand{\tabularnewline}{\\}
\providecolor{lyxadded}{rgb}{0,0,1}
\providecolor{lyxdeleted}{rgb}{1,0,0}

\usepackage{cite}

\renewcommand{\citedash}{\citeright\hbox{--}\penalty\@m\citeleft}
\renewcommand{\citepunct}{\citeright,\penalty\@m\hskip.13emplus.1emminus.1em\citeleft}

\makeatother

\usepackage{babel}

\makeatother

\usepackage{babel}
\begin{document}

\title{Time-Frequency Packing for High Capacity Coherent Optical Links}

\author{Giulio Colavolpe, \foreignlanguage{american}{\IEEEmembership{Senior Member, IEEE}},
and Tommaso Foggi{\normalsize{} }\\
\foreignlanguage{american}{\thanks{G. Colavolpe is with the University of Parma, Dipartimento di Ingegneria dell'Informazione, I-43124 Parma, Italy. T. Foggi is with the CNIT Research Unit at the University of Parma, I-43124 Parma, Italy.}\thanks{The paper was presented in part at the European Conference and Exhibition on Optical Communication (ECOC 2013), London, UK, September 2013, and at the IEEE Global Telecommunications Conference (GLOBECOM 2013), Atlanta, GA, USA, December 2013.}}}
\maketitle
\begin{abstract}
We consider realistic long-haul optical links, with linear and nonlinear
impairments, and investigate the application of time-frequency packing
with low-order constellations as a possible solution to increase the
spectral efficiency. A detailed comparison with available techniques
from the literature will be also performed. We will see that this
technique represents a feasible solution to overcome the relevant
theoretical and technological issues related to this spectral efficiency
increase and could be more effective than the simple adoption of high-order
modulation formats. \end{abstract}
\begin{IEEEkeywords}
Coherent optical systems, long-haul optical communications, faster-than-Nyquist
signaling, high-order modulation formats, spectral efficiency, time-frequency
packing.
\end{IEEEkeywords}

\section{Introduction}

The ever-growing bandwidth demand on internet and data networks has
pushed the research in the field of optical communications towards
more sophisticated transmission techniques~\cite{BigoOFC12,EssTka12,Liu:OFC11}.
This field has been characterized for decades by the simplest binary
formats and transmission systems, since bandwidth requirements were
met with both cost-effective and technological affordable solutions.
Nowadays optical communication channels have rapidly passed from 10
Gbps to 100 Gbps, whereas the next challenging step will lead to 1~Tbps.
Among the more severe limitations involved in such a system upgrade,
the technological and practical issues of processing high data rates
on a single channel, and the optical channel impairments related to
the required transmit power (i.e., the nonlinear effects) are the
most prominent. Since a higher capacity will hardly be reached with
single-channel transmissions, many different solutions and related
implementation techniques have been devised and proposed, in order
to exploit at best the optical channel capacity under current technological
costraints. Basically, all the proposed solutions, irrespectively
of the particular transmission technique, are based on multi-carrier
transmissions or so called \textquotedblleft{}superchannels\textquotedblright{}\cite{JansOFC12},
which means that the goal capacity is reached by binding up as many
single-channels together as necessary, in an efficient way. The techniques
investigated in this paper belong to this class and allow an effective
and reduced-complexity \textquotedblleft{}packing\textquotedblright{}
of the channels aiming at greatly improving the spectral efficiency
(SE) of the transmission.

As in most digital communication systems, orthogonal signaling is
the paradigm traditionally employed also in the design of long-haul
optical systems. This paradigm consists of ensuring the absence of
intersymbol interference (ISI) and, in multi-carrier scenarios, also
the absence of inter-carrier interference (ICI), i.e., the absence
of interference between adjacent channels. As an example, in single-carrier
coherent optical systems, possibly employing polarization multiplexing
(PM), given a conventional transmitter, with Mach-Zehnder (MZ) modulators
and return to zero (RZ) or non-return to zero (NRZ) shaping pulses,
when group velocity dispersion (GVD) and polarization mode dispersion
(PMD) are effectively compensated for and nonlinear effects are limited,
proper filtering and sampling at the receiver ensure that even a symbol-by-symbol
detector enables an almost-optimal performance since ISI and ICI are
very limited \cite{CoFoFoPr:JLT09}. Examples of multi-carrier transmission
systems based on this paradigm are represented by orthogonal frequency-division
multiplexing (OFDM) \cite{ZaEl11} (see also \cite{BaCoFoFoPr:JLT10}
and references therein) and Nyquist wavelength-division multiplexing
(WDM) \cite{BoCuCaPoFo11} (and similar bandwidth narrowing techniques
such as \cite{LiTipErKarlAnd:JLT12}), where the use of proper shaping
pulses allows to remove, at least in theory, both ISI and ICI without
using guard bands and thus without wasting resources. On the other
hand, when practical transmit or receive filters are considered and
nonlinear effects come into play, orthogonality is no more guaranteed
and an unwanted interference appears. In addition, in these orthogonal
signaling systems, the spectral efficiency can be improved only by
increasing the constellation cardinality, thus employing modulation
formats that are more sensitive to nonlinear effects and crosstalk.

An alternative paradigm to increase the SE is represented by time-frequency
packing (TFP) \cite{BaFeCo09b,CoFoMoPi11,CaDaLu12,PiMoCoAl13}. In
this case, low-order modulations, such as quaternary phase shift keying
(QPSK), are employed but the spacing between two adjacent pulses in
the time domain (i.e., the symbol interval) is reduced well below
that corresponding to the Nyquist rate. Similarly, the frequency separation
between two adjacent channels can be also reduced, with the aim of
maximizing the \emph{achievable spectral efficiency}, which is thus
used as a performance measure instead of the minimum Euclidean distance
(as in more classical \emph{faster-than-Nyquist} signaling schemes,
see \cite{CoFoMoPi11} and references therein). In addition, rather
than the optimal receiver, a complexity-constrained detector is considered.
In other words, controlled ISI and ICI are introduced and partially
coped with at the receiver either through a single-channel maximum
a posteriori (MAP) symbol detector, which is designed to take into
account only a limited amount of ISI, or through a suboptimal multiuser
detector (MUD), which enables the joint processing of multiple sub-channels
and thus coping with a limited amount of ICI (and possibly ISI). 

Advanced signal processing plays a key role in TFP systems since,
besides the fundamental working principles which entail decoding and
soft detection techniques, significant performance improvements derive
from a skillful combination of pulse shaping, coding, the adoption
of a proper linear filtering (the channel shortening technique described
in \cite{RuPr12}), of a proper trellis-based MAP symbol detection
strategy, and possibly of a proper multiuser processing. The aim is
the maximization of the achievable SE, computed by resorting to the
simulation-based method described in \cite{ArLoVoKaZe06} which allows
to also take nonlinear effects into account since it holds for \textbf{any}
channel, including nonlinear and non-Gaussian. 

With respect to \cite{BaFeCo09b}, which considers the case of an
additive white Gaussian noise (AWGN) channel and a simple symbol-by-symbol
detector after matched filtering, and\textbf{ }\cite{CoFoMoPi11}
which considers the optical channel in the linear regime and a more
sophisticated receiver based on trellis processing, in this paper
we have the following novel main contributions. (i) First, we compute
the spectral efficiency and optimize the system parameters \emph{by
taking into account nonlinear effects}. (ii) Then, the considered
receiver structures are different. We consider a trellis-based receiver
enhanced by the use of the channel shortening technique \cite{RuPr12}.
This makes a big difference in terms of performance. We also consider
here the case of use of a nonlinear compensation technique at the
receiver based on digital backpropagation. (iii) Finally, we compare
the performance of the TFP technique with other solutions in the literature.
A similar investigation has been performed in \cite{PiMoCoAl13} with
reference to the nonlinear satellite channel. Satellite nonlinearities
are, in nature, very different from those affecting long-haul optical
transmissions. The results and the conclusions here reported are thus
very different from those reported in \cite{PiMoCoAl13}. In particular,
we will see here that in long-haul optical links with strong nonlinear
effects, an increase of SE cannot be simply obtained by increasing
the modulation order.

The remainder of this paper is organized as follows. The system model
is described in Section \ref{sec:System-Model}. The framework that
we use to evaluate the SE is then described in Section \ref{sec:Spectral-Efficiency-optimization},
whereas the adopted receivers are described in Section \ref{sec:Auxiliary-Channel-Model}.
Numerical results are reported in Section \ref{sec:Simulation-Results}
with a detailed comparison with alternative approaches in the literature,
possibly based on the paradigm of orthogonal signaling. Finally, some
conclusions are drawn in Section \ref{sec:Conclusion}.

\section{Preliminaries and System Model\label{sec:System-Model}}

Let us first consider a multi-carrier system over an AWGN channel,
where $N_{c}$ equally-spaced adjacent carriers are associated to
the same linear modulation format and shaping pulse $p(t)$. The complex
envelope of the transmitted signal can be expressed as%
\footnote{In the following, we will consider the adoption of polarization multiplexing.
In this case, $s(t)$ is the signal transmitted on one state of polarization.%
}

\begin{equation}
s(t)=\sum_{\ell}\sum_{k=0}^{K-1}x_{k}^{(\ell)}p(t-kT-\tau^{(\ell)})e^{j(2\pi\ell Ft+\theta^{(\ell)})}\label{eq:rec_signal}
\end{equation}
where $K$ is the number of symbols transmitted over each carrier,
$T$ the symbol interval, $x_{k}^{(\ell)}$ the symbol transmitted
over the $\ell$th carrier during the $k$th symbol interval, $\tau^{(\ell)}$
and $\theta^{(\ell)}$ the delay and the initial phase of the $\ell$th
carrier, respectively, and $F$ the frequency spacing between two
adjacent carriers.

In such a scenario, the capacity-achieving distribution of the transmitted
symbols is Gaussian. Hence, in order to maximize the system spectral
efficiency, independent and uniformly distributed (i.u.d.) Gaussian
symbols $\{x_{k}^{(\ell)}\}$ must be employed along with a shaping
pulse $p(t)$ having a rectangular spectrum (sinc pulse) with bandwidth
$B=1/2T$ and a frequency spacing $F=1/T$, i.e., no guard band is
employed between two adjacent carriers. No ISI or ICI occur since
orthogonal signaling is employed. 

Practical systems necessarily deviate from this paradigm. First of
all, the transmitted symbols $\{x_{k}^{(\ell)}\}$ are not Gaussian
but usually belong to a properly normalized zero-mean $M$-ary complex
constellation $\chi$. Under these conditions, instead of trying to
approach as close as possible the impractical condition of having
a shaping pulse with rectangular spectrum and to reduce as much as
possible the guard band, as in Nyquist-WDM systems \cite{BoCuCaPoFo11},
TFP technique intentionally introduces both ISI and ICI to improve
the spectral efficiency \cite{BaFeCo09b,CoFoMoPi11}. In other words,
for a given shaping pulse, the symbol time $T$ and the frequency
spacing $F$ are properly optimized, as explained in Section~\ref{sec:Spectral-Efficiency-optimization},
to maximize the spectral efficiency by intentionally violating the
orthogonal signaling paradigm. 

\begin{figure*}
\begin{centering}
\includegraphics[width=0.8\paperwidth]{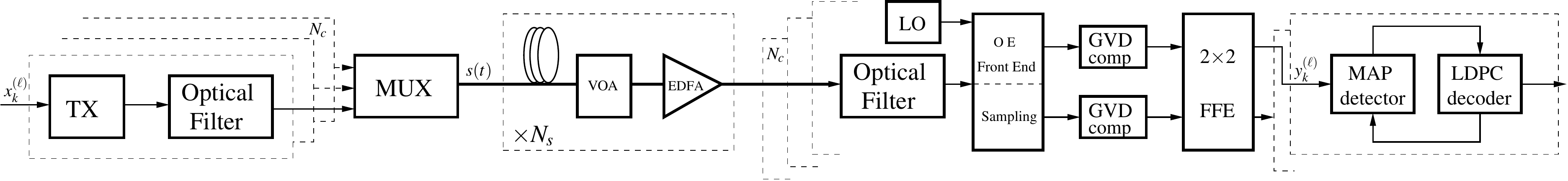}
\par\end{centering}

\caption{\label{fig:system-model}Schematic of the transmission system, where
$N_{s}$ is the number of link spans, $N_{c}$ the number of carriers,
EDFA an \textit{erbium-doped fiber amplifier}, LO a \textit{local
oscillator}, O/E Front End the \emph{opto-electronic front end}~\cite{CoFoFoPr:JLT09},
GVD comp. a \emph{linear equalizer} aimed at compensating for the
chromatic dispersion, the $2\times2$ DD-FFE is a two-dimensional
\textit{decision-directed feed-forward equalizer,} which is then followed
by iterative detection/decoding between a proper detector and the
low-density parity-check (LDPC) decoder.}
\end{figure*}

Let us now consider a realistic optical system. In this case, the
possibility to generate a transmitted signal with expression~(\ref{eq:rec_signal})
is strictly related to the availability of a linear modulator. In
other words, let us consider the transmitted signal associated to
the carrier for $\ell=0$. If the pulse $p(t)$ has support larger
than $T$, this signal cannot be directly generated through an MZ
modulator unless it is properly linearized. This is due to the nonlinear
transfer function of the MZ modulator between the electrical signal
at its input and the optical signal at its output. We could, however,
use an MZ modulator to generate a linearly modulated signal with shaping
pulse having support at most $T$ and then ``stretch'' the transmitted
pulses through an optical filter, thus obtaining an effect similar
to that obtained with time packing. Hence, in this case, time-packing
is not an available option but we have a viable surrogate. The degrees
of freedom are thus the frequency spacing $F$ and the bandwidth $B$
of the optical filter used at the MZ output~\cite{CoFoMoPi11}, which
in the present analysis is always a 4th-order Gaussian filter (both
at transmit and receive side). Fig.~\ref{fig:system-model} shows
a schematic of the considered transmission system, irrespectively
of the constellation size. Blocks related to the receiver will be
explained in Sections \ref{sec:Auxiliary-Channel-Model} and \ref{sec:Simulation-Results}.

We consider optical channels impaired by GVD and PMD, and in particular
uncompensated links where chromatic dispersion compensation is only
performed with a fixed-tap equalizer in the electrical domain at the
receive side. We also take into account nonlinear effects, as it will
be explained in Section~\ref{sec:Simulation-Results}, where a description
of the simulated optical links will be provided. We will compare systems
based on TFP and employing quaternary constellations with other known
systems, based on higher-order modulations, which show good results
in terms of spectral efficiency, namely Nyquist-WDM systems \cite{BoCuCaPoFo11}
and the receiver-side duobinary shaping in \cite{LiTipErKarlAnd:JLT12}.
In the latter system, an electrical two-tap filter is used to force
a duobinary shape to the received signal, thus limiting the ICI, at
the expense of employing a MAP sequence (or symbol) detector to cope
with the introduced ISI. This is, in practice, a heuristic version
of frequency packing and the processing here described for TFP systems
since it allows a larger packing in frequency at the transmitter.
No comparison will be performed with OFDM systems since, from an implementation
point of view, they present a few drawbacks for optical links and
are also less efficient than the systems considered here~\cite{BaCoFoFoPr:JLT10}.

\section{Spectral Efficiency Computation\label{sec:Spectral-Efficiency-optimization}}

We now describe the framework used to evaluate the performance limits
of all optical transmission systems considered in this paper and to
perform the optimization of the optical filter bandwidth and frequency
spacing in case of adoption of the TFP technique. 

We are considering an optical channel with linear and nonlinear distortions,
simulated through the split-step Fourier method (SSFM) with proper
complexity \cite{Men03}. Denoting by $\mathbf{y}$ a proper discrete-time
received sequence used for detection of the information symbols $\mathbf{x}=\{x_{k}^{(\ell)}\}_{k,\ell}$,
the information rate (IR), i.e., the average mutual information when
the information symbols are independent and uniformly distributed
(i.u.d.) random variables belonging to the given constellation, is
defined as 
\begin{equation}
I(\mathbf{x};\mathbf{y})\!=\!\!\!\lim_{K\rightarrow\infty}\frac{1}{N_{c}K}E\Bigg\{\log_{2}\frac{p(\mathbf{y}\arrowvert\mathbf{x})}{\sum_{\mathbf{x'}}p(\mathbf{y}\arrowvert\mathbf{x'})P(\mathbf{x'})}\Bigg\}\,\,\left[\frac{\mbox{bit}}{\mbox{ch. use}}\right]\label{e:IR-1}
\end{equation}
where $p(\cdot)$ denotes a probability density function (PDF) and
$P(\cdot)$ a probability mass function (PMF). In (\ref{e:IR-1}),
the information symbols are, as mentioned, i.u.d. and thus $P(x_{k}^{(\ell)})=1/M$,
for all $k,\ell$. The spectral efficiency (SE) is the IR per unit
bandwidth and unit time and reads 
\[
\frac{I(\mathbf{x};\mathbf{y})}{FT}\quad[\mbox{b/s/Hz}]
\]
since $FT$ is the time-frequency slot devoted to the transmission
of symbol $x_{k}^{(\ell)}$.

The computation of IR and SE requires the availability of the pdfs
$p(\mathbf{y}\arrowvert\mathbf{x})$ and $p(\mathbf{y})=\sum_{\mathbf{x'}}p(\mathbf{y}\arrowvert\mathbf{x'})P(\mathbf{x'})$.
However, they are not known in closed form nor can we resort to the
simulation method in \cite{ArLoVoKaZe06} to compute them. In fact,
this method requires that the channel at hand is finite-state and
the availability of an optimal detector for it \cite{ArLoVoKaZe06}.
These conditions are clearly not satisfied in our scenario \cite{EssTka12,Essjlt10}.
We may thus resort to the computation of a proper lower bound on the
IR (and thus on the SE) obtained by substituting $p(\mathbf{y}\arrowvert\mathbf{x})$
in (\ref{e:IR-1}) with an arbitrary auxiliary channel law $q(\mathbf{y}\arrowvert\mathbf{x})$
with the same input and output alphabets as the original channel (mismatched
detection~\cite{MeKaLaSh94,ArLoVoKaZe06,CoFoMoPi11,Serena12}). The
resulting lower bound reads
\begin{equation}
\underline{I}_{q}(\mathbf{x};\mathbf{y})\!=\!\!\!\lim_{K\rightarrow\infty}\frac{1}{N_{c}K}E\Bigg\{\log_{2}\frac{q(\mathbf{y}\arrowvert\mathbf{x})}{\sum_{\mathbf{x'}}q(\mathbf{y}\arrowvert\mathbf{x'})P(\mathbf{x'})}\Bigg\}\,\,\left[\frac{\mbox{bit}}{\mbox{ch. use}}\right].\label{e:LB}
\end{equation}
If the auxiliary channel law is representative of a finite-state channel,
pdfs $q(\mathbf{y}\arrowvert\mathbf{x})$ and $q_{p}(\mathbf{y})=\sum_{\mathbf{x'}}q(\mathbf{y}\arrowvert\mathbf{x'})P(\mathbf{x'})$
can be computed, this time, by using the optimal MAP symbol detector
for that auxiliary channel \cite{ArLoVoKaZe06}. This detector, that
will be clearly suboptimal for the actual channel, will have at its
input the sequence $\mathbf{y}$ generated by simulation \emph{according
to the actual channel model,} and the expectation in\emph{ }(\ref{e:LB})
is meant with respect to the input and output sequences generated
accordingly \cite{ArLoVoKaZe06}. Thus, no assumption on the real
statistics of the discrete-time received sequence is required for
the design of the adopted detector since it is designed for the auxiliary
channel. Similarly, the knowledge of the real statistics of the sequence
$\mathbf{y}$ are not required for its generation since it can be
obtained by simulation through the SSFM. If we change the adopted
receiver (or, equivalently, if we change the auxiliary channel) we
obtain different lower bounds on the information rate but, in any
case, these bounds are \emph{achievable} by those receivers, according
to mismatched detection~\cite{MeKaLaSh94,ArLoVoKaZe06}. We will
thus say, with an abuse of terminology, that the computed lower bounds
are the SE values of the considered channel when those receivers are
employed. All these considerations hold for any actual channel including
nonlinear and non-Gaussian ones.

This technique thus allows to take into account receivers with reduced
complexity. In fact, it is sufficient to consider an auxiliary channel
which is a simplified version of the actual channel in the sense that
only a portion of the actual channel memory and/or a limited number
of impairments are present. The considered receivers will be described
in the next section. However, we may anticipate that for all of them
we will assume that parallel independent detectors are employed, one
for each carrier (and each polarization in case of polarization multiplexing).
In other words, ICI is not coped with at the receiver since multiuser
detection is considered too computationally demanding. This corresponds
to the adoption of an auxiliary channel model that can be factorized
into the product
\[
q(\mathbf{y}\arrowvert\mathbf{x})=\prod_{\ell}q(\mathbf{y}^{(\ell)}\arrowvert\mathbf{x}^{(\ell)})
\]
where $\mathbf{y}^{(\ell)}$ is a proper discrete-time received sequence
used for detection of symbols $\mathbf{x}^{(\ell)}=\{x_{k}^{(\ell)}\}$
transmitted over the $\ell$th carrier. Under this hypothesis and
assuming a system with a large number of carriers in order to neglect
border effects, it simply results 
\begin{equation}
\underline{I}_{q}(\mathbf{x};\mathbf{y})=\lim_{K\rightarrow\infty}\frac{1}{K}E\Bigg\{\log_{2}\frac{q(\mathbf{y}^{(\ell)}\arrowvert\mathbf{x}^{(\ell)})}{q_{p}(\mathbf{y}^{(\ell)})}\Bigg\}\,,\label{e:IR-2}
\end{equation}
i.e., the result can be computed by considering only one carrier and
does not depend on the specific considered carrier. In a practical
scenario with a finite number of carriers, we will consider the central
carrier only, avoiding the computation on the border carriers which
are affected by a lower amount of ICI, thus obtaining a further lower
bound. Without loss of generality, we will assume that the central
carrier is that with $\ell=0$.

Note that, as stated, we are not able to compute the IR of the actual
channel, but this is irrelevant because the optimal receiver for the
actual optical channel is unavailable and thus we can in no way achieve
it. The best we can do is to employ practical suboptimal receivers
and for them we are indeed able to compute the relevant IR which will
be called achievable IR. The corresponding achievable (lower bound
on) SE is thus 
\begin{equation}
\underline{\eta}=\frac{1}{FT}\underline{I}_{q}(\mathbf{x};\mathbf{y})\quad[\mbox{\textrm{b/s}/\textrm{Hz}}].\label{e:eta}
\end{equation}
The aim of the TFP technique is to find the values of $F$ and the
bandwidth $B$ of the optical filter after the MZ modulator providing,
for each value of the signal-to-noise ratio (SNR) or, equivalently,
for each value of the transmitted power, the maximum value of SE achievable
by that particular receiver, which is optimal for the considered specific
auxiliary channel. Namely, we compute 
\begin{equation}
\underline{\eta}_{\text{M}}=\max_{F,B>0}\underline{\eta}(F,B)\,.\label{e:eta_M}
\end{equation}
Typically, the dependence on the SNR value is not critical, in the
sense that we can identify two or at most three SNR regions for which
the optimal spacings practically have the same value.

For fair comparisons in terms of SE, we need a proper definition of
the SNR. We define the SNR as the ratio $P/N$ between the signal
power and the noise power (in the considered bandwidth). Under the
assumption of a large number of carriers to avoid boundary effects,
$P/N$ can be written as 
\begin{equation}
\frac{P}{N}=\lim_{N_{c}\rightarrow\infty}\frac{N_{c}P_{c}}{B_{o}2N_{0}}\label{e:SNR}
\end{equation}
where $P_{c}$ is the power for each carrier, $B_{o}$ the overall
bandwidth, and $N_{0}/2$ the two-sided power spectral density of
the amplified spontaneous emission (ASE) noise per polarization, as
if the channel were linear. $P_{c}$ is independent of the bandwidth
$B$. It is clearly $B_{o}=(N_{c}-1)F+B$. In the limit of a large
number of carriers, i.e., when border effects can be neglected, or
when $B$ is comparable with $F$, we may approximate $B_{o}\simeq N_{c}F$
and thus
\begin{equation}
\frac{P}{N}\simeq\frac{P_{c}}{2N_{0}F}\,.\label{e:SNR2}
\end{equation}
The SNR definition as given in (\ref{e:SNR2}) is independent of the
transmit waveform and its parameters. This definition will be adopted
even when the number of carriers $N_{c}$ is rather small and, if
we neglect the border effects, corresponds to the SNR per carrier.
This provides a common measure to compare the performance of different
solutions in a fair manner and allows also to compare the SE values
obtained under nonlinear propagation conditions with the Shannon limit
for the bandlimited AWGN channel. This will allow to appreciate the
degradation due to the nonlinear effects.

\section{Considered Receivers\label{sec:Auxiliary-Channel-Model}}

The system model described in Section~\ref{sec:System-Model} is
representative of the considered scenario and has been employed in
the information-theoretic analysis and in the simulations results.
In this section, we describe two families of employed receivers and
the corresponding auxiliary channels (the channels for which those
receivers represent the optimal MAP symbol detectors).%
\footnote{In each family, we have one receiver for each considered transmission
system, namely TFP, Nyquist-WDM, and receiver-side duobinary shaping.%
}

The first family is composed of receivers which completely neglect
nonlinear distortions. Hence, the corresponding auxiliary channels
operate in the linear regime. As far as GVD and PMD are concerned,
they are assumed perfectly compensated. As known, in the absence of
nonlinear effects perfect compensation is possible through a proper
two-dimensional equalizer~\cite{CoFoFoPr:JLT09}.%
\footnote{In Fig. \ref{fig:system-model}, this equalizer is represented as
the cascade of two fixed one-dimensional equalizers, one for each
polarization, aimed at compensating for the GVD, and a short adaptive
two-dimensional equalizer to cope with the PMD. %
} We also mentioned that multiuser detection is not considered at the
receiver, i.e., in the considered auxiliary channels ICI is also neglected.
Under these assumptions, the independent detectors mentioned in the
previous section, one for each carrier and each polarization, have
to take into account only (a portion of) the ISI intentionally or
accidentally introduced in the system. As an example, in the case
of TFP or the adoption of the receiver-side duobinary shaping \cite{LiTipErKarlAnd:JLT12},
the detector takes into account (a portion of) the ISI intentionally
introduced. In the case of Nyquist-WDM systems, a receiver coping
with the unwanted ISI deriving from the adopted practical filters
and shaping pulses is considered instead.

In the second family of auxiliary channels, part of the nonlinear
effects are compensated through digital backpropagation \cite{Essjlt10}.
The remaining nonlinear effects\emph{ }(such as signal-ASE noise interaction
that cannot be compensated by digital backpropagation)\emph{ }are
neglected. Then we proceed as in the previous case. So, in practice,
after the possible digital backpropagation and the two-dimensional
equalizer, the auxiliary channels are the same in both cases. The
expression of $q(\mathbf{y}^{(0)}|\mathbf{x}^{(0)})$ (remember that
we are considering the central carrier only) will be provided at the
end of the section.

In practice, we are computing the SE achievable on the optical channel
when two possible receiver designs are adopted. In the first one,
nonlinear effects are neglected at the receiver (i.e., the receiver
is designed for the linear regime). This is obviously a worst case.
The second case is when we adopt the best available technique for
the compensation of nonlinear effects. As shown in Section \ref{sec:Simulation-Results},
the presence or absence of digital backpropagation only affects the
maximum values of achievable SE but not significantly our conclusions.
We expect that they will hold also when we employ other compensation
techniques. 

As mentioned, for the TFP technique the carrier spacing and the bandwidth
of the transmit optical filter are optimized to maximize the spectral
efficiency. For a fair comparison, in the case of Nyquist-WDM systems
we also optimize, from an SE point of view, the transmit optical filter
bandwidth and the channel spacing, with the constraint that the spacing
is not smaller than the Nyquist bandwidth, i.e., $1/T$ (otherwise
we fall in the domain of the TFP technique). Under these conditions
the memory of the channel in the absence of nonlinear effects is usually
limited to at most $L=2$ interfering symbols (it can be simply verified
by increasing the value of $L$ assumed at the receiver and noting
that no improvement is attained). On the other hand, the memory at
the receiver in the case of the technique described in \cite{LiTipErKarlAnd:JLT12}
is, by definition, $L=1$. The memory introduced by the TFP technique
is, instead, potentially very large. To limit the receiver complexity
with a limited performance degradation, we apply a channel shortening
(CS) technique~\cite{FaMa73}. In fact, when the memory of the channel
is too large to be taken into account by a full complexity detector,
an excellent performance can be achieved by properly filtering the
received signal before adopting a reduced-state detector~\cite{FaMa73}.
A very effective CS technique for general linear channels is described
in~\cite{RuPr12}. 

In the case of adoption of CS, we are looking for an auxiliary channel
and the corresponding optimal MAP symbol detector. As mentioned, in
the auxiliary channel model we consider that nonlinearities are absent
or perfectly compensated whereas GVD and PMD are assumed perfectly
compensated. In addition, independent receivers, one for each carrier
and each polarization are considered here since, as mentioned, ICI
is neglected in the auxiliary channel model. Hence, each receiver
assumes that, apart from AWGN, only one carrier is present. A set
of sufficient statistics $\mathbf{y}^{(0)}$ can thus be obtained
by sampling the output of a filter matched to the transmit pulse (matched
filter, MF). The $k$th element of $\mathbf{y}^{(0)}$, under the
above mentioned assumption that only the carrier with $\ell=0$ is
present, reads 
\[
y_{k}^{(0)}=\sum_{i}x_{k-i}^{(0)}g_{i}+n_{k}
\]
where 
\[
g_{i}=\int h(t)h^{*}(t-iT)\mathrm{d}t\,,
\]
$h(t)$ being the convolution of the shaping pulse $p(t)$ after the
MZ modulator and the optical transmit filter impulse response, and
$n_{k}$ a Gaussian process with $\mathrm{E}\{n_{k+i}n_{k}^{*}\}=2N_{0}g_{i}$.
Vector $\mathbf{y}^{(0)}$ can be written as 
\begin{equation}
\mathbf{y}^{(0)}=\mathbf{G}\mathbf{x}^{(0)}+\mathbf{n}\label{e:sufficient_statistics}
\end{equation}
where $\mathbf{G}$ is a Toeplitz matrix obtained from the sequence
\{$g_{i}$\} whereas $\mathbf{n}$ is a vector collecting the colored
noise samples. This is the so called \emph{Ungerboeck observation
model} for ISI channels~\cite{CoBa05b}. According to the CS approach,
the considered auxiliary channel is based on the following channel
law~\cite{RuPr12} 
\begin{equation}
q(\mathbf{y}^{(0)}|\mathbf{x}^{(0)})\propto\exp\left(2\Re(\mathbf{y}^{(0)}{}^{H}\mathbf{H}^{r}\mathbf{x}^{(0)})-\mathbf{x}^{(0)}{}^{H}\mathbf{G}^{r}\mathbf{x}^{(0)}\right)\,,\label{e:mis_channel_law}
\end{equation}
where $(\cdot)^{H}$ denotes transpose conjugate, whereas $\mathbf{H}^{r}$
and $\mathbf{G}^{r}$ are Toeplitz matrices obtained from proper sequences
\{$h_{i}^{r}$\} and \{$g_{i}^{r}$\}, and are known as $\textit{channel shortener}$
and $\textit{target response}$, respectively~\cite{RuPr12}. Matrix
$\mathbf{H}^{r}$ represents a linear filtering of the sufficient
statistics~($\ref{e:sufficient_statistics}$), and $\mathbf{G}^{r}$
is the ISI to be set at the detector (different from the actual ISI)~\cite{RuPr12}.
In~($\ref{e:mis_channel_law}$), the noise variance has been absorbed
into the two matrices. In order to reduce the complexity, we constrain
the target response used at the receiver to 
\begin{equation}
g_{i}^{r}=0\quad|i|>L_{r}\,,\label{e:G_constraint_psk}
\end{equation}
which implies that the memory of the detector is $L_{r}$ instead
of the true memory $L$ of the channel. The CS technique finds a closed
form of the optimal \{$h_{i}^{r}$\} and \{$g_{i}^{r}$\} which maximize
the achievable IR~($\ref{e:IR-2}$).%
\footnote{This closed-form expression is derived under the assumpion of Gaussian
input symbols. However, by using these filter and target response
in the presence of symbols belonging to finite constellations, an
impressive performance improvement is still observed~\cite{RuPr12}.%
} If the memory $L_{r}$ is larger than or equal to the actual channel
memory, as in the case of Nyquist-WDM or receiver-side duobinary shaping,
the trivial solution is $\mathbf{H}^{r}=\mathbf{I}/2N_{0}$ and $\mathbf{G}^{r}=\mathbf{G}/2N_{0}$,
where $\mathbf{I}$ is the identity matrix. Interestingly, when $L_{r}=0$
the optimal channel shortener becomes a minimum mean square error
(MMSE) feedforward equalizer~\cite{RuPr12}. The optimal MAP symbol
detector for the auxiliary channel with law~(\ref{e:mis_channel_law}),
that we used for the computation of the achievable IR according to
the technique in \cite{ArLoVoKaZe06}, is described in~\cite{CoBa05b}
(see also \cite{RuPr12,RuFe12} for further details).

\section{Numerical Results\label{sec:Simulation-Results}}

\selectlanguage{american}%
\begin{table*}[t]
\selectlanguage{english}%
\caption{\selectlanguage{american}%
\label{tab:NL-link}\foreignlanguage{english}{SMF lengths for each
span in the simulated optical link.}\selectlanguage{english}%
}

\centering{}%
\begin{tabular}[b]{|c||c|c|c|c|c|c|c|c|c|c|c|c|c|c|c|}
\hline 
span \# & 1 & 2 & 3 & 4 & 5 & 6 & 7 & 8 & 9 & 10 & 11 & 12 & 13 & 14 & 15\tabularnewline
\hline 
SMF (km) & 70.8 & 75.5 & 55.1 & 52.1 & 40.1 & 67. & 53.2 & 50 & 80.3 & 79.1 & 53.6 & 75.1 & 90.3 & 54.2 & 99.4\tabularnewline
\hline 
\end{tabular}\selectlanguage{american}%
\end{table*}
\foreignlanguage{english}{We here report the maximum achievable spectral
efficiency $\underline{\eta}_{\textrm{M}}$ as a function of $P/N$
for different polarization-multiplexed systems. We assume perfect
synchronization as we are interested in the evaluation of achievable
bounds for the spectral efficiency. Unless otherwise stated, the employed
shaping pulses are those resulting from the use of RZ pulses with
duty cycle 50\% , an MZ modulator, and a 4th-order Gaussian optical
transmit filter with 3-dB optical filter bandwidth~$B$ (specified
later in this section). The considered modulation formats are QPSK
and 16/64/256-ary quadrature amplitude modulations (QAMs). We first
considered systems with 8 carriers (subchannels) at 140 Gbps each,
irrespectively of the employed modulation format (thus, the baud rate
changes for each format). It can be noticed that, since the bandwidth
of each subchannel is highly reduced by filtering and multilevel modulations
are considered, the required sampling rate is always within the state-of-the-art
technology, i.e, well below $80$ Gsample/s. At the receiver side,
after the 4th-order Gaussian optical receive filter, with 3-dB optical
filter bandwidth $B_{R}$, opto-electric conversion through an optical
hybrid, and sampling, two non-adaptive one-dimensional equalizers
(one for each polarization) perform coarse GVD compensation. Then,
a two-dimensional (2-D) decision-directed (DD) adaptive feedforward
equalizer (FFE) with 25 taps process the signals received over two
orthogonal states of polarization to compensate for residual GVD,
to demultiplex polarizations,}%
\footnote{\selectlanguage{english}%
Note that in the simulated scenarios no PMD is present. We also performed
simulations including PMD for the link in Table \ref{tab:NL-link},
for which we had at our disposal the measured differential group delay,
but no difference has been noticed. \selectlanguage{american}%
}\foreignlanguage{english}{ and to complete (along with the optical
filter) the implementation of the MF~\cite{CoFoFoPr:JLT09}.}%
\footnote{\selectlanguage{english}%
In other words, the FFE taps are updated using the MF output as target
response, so that the equalizer does not remove the ISI induced by
narrow filtering. It is worth noting that, if extremely narrow optical
filtering is employed at the receive side, the electrical compensation
of chromatic dispersion through the non-adaptive equalizers may result
to be inaccurate. In this case, a wider optical filter can be used,
compatibly with the system design, in order to leave the useful component
of the received signal unchanged, whereas matched filtering is implemented
by the adaptive equalizer. \selectlanguage{american}%
}\foreignlanguage{english}{ The number of taps of the two non-adaptive
one-dimensional equalizers and that of the 2-D DD adaptive feedforward
equalizer have been selected in such a way they do not affect the
performance. In other words, no performance improvement has been observed
by increasing the number of taps. In the case of systems employing
receiver-side duobinary shaping, a further digital filter is present
to perform the required shaping~\cite{LiTipErKarlAnd:JLT12}. The
output is provided to the MAP symbol detectors (one for each carrier
and each polarization) which iteratively exchange information with
the decoders for a maximum of 50 iterations.}
\begin{figure}
\selectlanguage{english}%
\begin{centering}
\includegraphics[width=1\columnwidth]{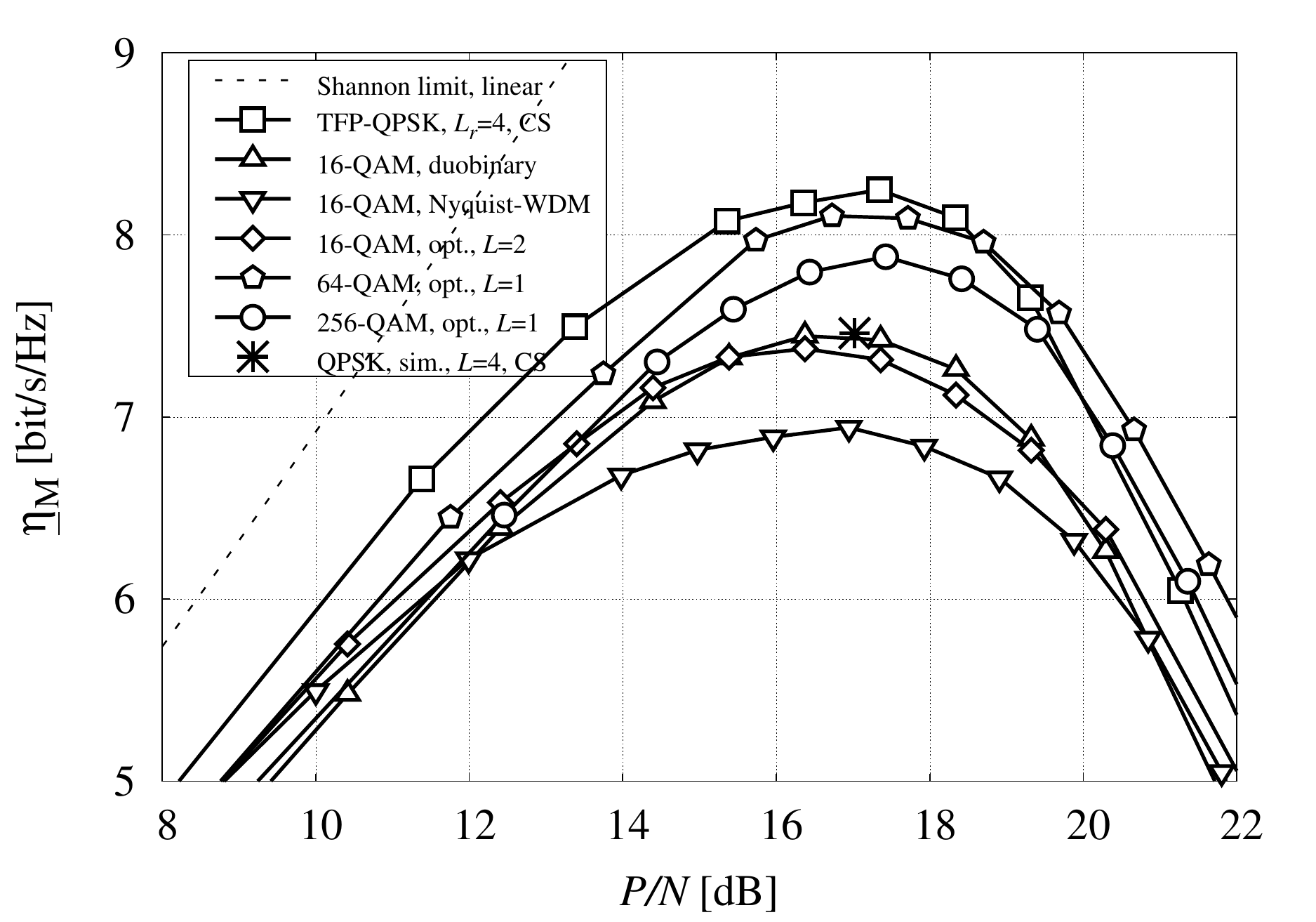}
\par\end{centering}

\selectlanguage{american}%
\caption{\label{fig:SE1}\foreignlanguage{english}{Maximum achievable spectral
efficiency on the considered uncompensated link, for TFP-QPSK, 16-QAM
with Nyquist-WDM spacing, 16-QAM with receiver-side duobinary shaping,
and SE-optimized 16/64/256-QAM, all systems with 8 140-Gbit/s sub-channels.
The Shannon limit in the linear regime is also reported as a reference,
along with the simulated performance of TFP-QPSK employing a rate-4/5
LDPC code.}}
\end{figure}

\begin{figure}
\selectlanguage{english}%
\begin{centering}
\includegraphics[width=1\columnwidth]{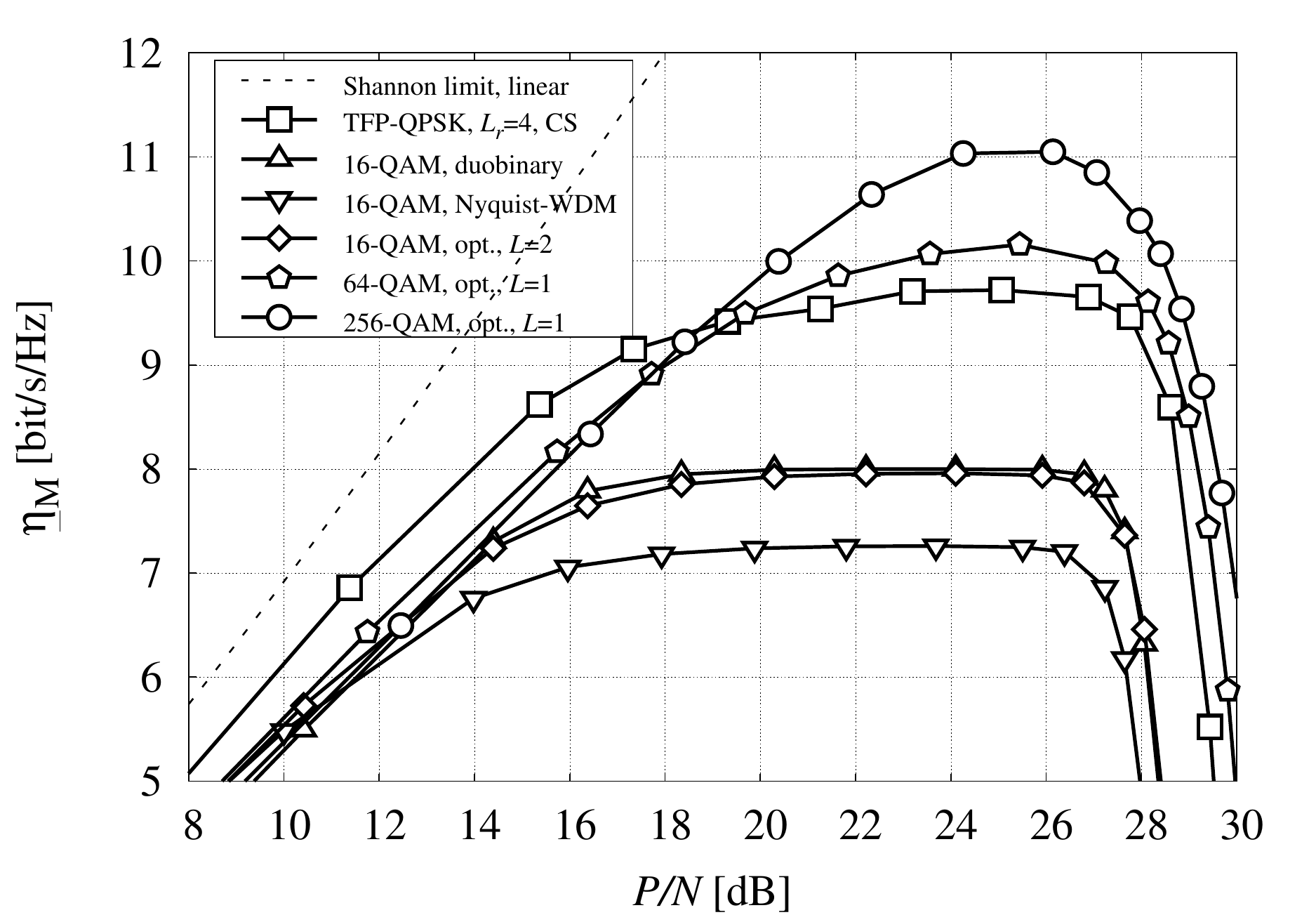}
\par\end{centering}

\selectlanguage{american}%
\caption{\selectlanguage{english}%
\label{fig:SE2}Maximum achievable spectral efficiency on the considered
uncompensated link for TFP-QPSK and $M$-QAM of Fig.~\ref{fig:SE1}
with ideal digital backpropagation.\selectlanguage{american}%
}
\end{figure}

\selectlanguage{english}%
All the considered constellations can be viewed, with a proper rotation,
as two independent signals transmitted over the in-phase and quadrature
components, respectively. Hence, at the receiver side, we may use
two identical and independent detectors, one working on the in-phase
and the other one on the quadrature component. This is beneficial
in case of adoption of a MAP detector. In fact, when $L$ interfering
symbols are taken into account, we have two detectors (per polarization)
working on a trellis with $(\sqrt{M})^{L}$ states instead of a single
detector working on a trellis with $M^{L}$ states. Hence, for a given
complexity, a larger memory can be taken into account. 

All spectral efficiency computations have been performed, for each
point of the presented curves, on pseudo-random sequences of 900,000
bits per quadrature, following a training phase of 100,000 bits. Each
sequence has been split into blocks of 50,000 bits that, properly
processed as described in \cite{ArLoVoKaZe06}, allow us to compute
the achievable IR. The confidence interval on the computed information
rate turned out to be less than 2\%. 
\begin{table}
\caption{\label{tab:Bandwidths}Normalized bandwidths for transmit- and receive-side
optical filters. ``Variable'' means that the bandwidth depends on
the considered SNR.}

\centering{}%
\begin{tabular}{|c|c|c|}
\hline 
 & $B$ & $B_{R}$\tabularnewline
\hline 
TFP-QPSK & variable & variable\tabularnewline
\hline 
16-QAM \cite{BoCuCaPoFo11} & $1.1/T$ & $1/T$\tabularnewline
\hline 
16-QAM \cite{LiTipErKarlAnd:JLT12} & $1/T$ & $1/T$\tabularnewline
\hline 
16-QAM & $1/T$ & $0.7/T$\tabularnewline
\hline 
64-QAM & $1.1/T$ & $0.85/T$\tabularnewline
\hline 
256-QAM & $1.25/T$ & $0.9/T$\tabularnewline
\hline 
\end{tabular}
\end{table}

We considered an existing link of standard single-mode fiber (SMF),
whose spans have the lengths reported in Table~\ref{tab:NL-link},
simulated through the SSFM. The fiber dispersion is 16.63 $\mathrm{ps/nm/km}$,
the attenuation is 0.23 $\mathrm{dB/km}$, the nonlinear index $\gamma$
is equal to 1.3 $\mathrm{W^{-1}km^{-1}}$, and the noise figure of
all amplifiers is equal to 6 $\mathrm{dB}$. For this link, Fig.~\ref{fig:SE1}
shows the maximum achievable spectral efficiency $\underline{\eta}_{\text{M}}$
for a TFP system employing QPSK (TFP-QPSK), a Nyquist-WDM system using
a 16-QAM, a system employing receiver-side duobinary shaping, still
with 16-QAM, and SE-optimized 16/64/256-QAM systems. In the case of
TFP, the frequency spacing $F$ and the optical 3-dB filter bandwidth
$B$ and $B_{R}$ have been optimized for each value of $P/N$, resulting
in \textbf{$B_{R}=B$}. For the Nyquist-WDM systems, we used $B=1.1/T$,
$B_{R}=1/T$, and an NRZ shaping pulse before the MZM, as suggested
in \cite{BoCuCaPoFo11} whereas, for receiver-side duobinary shaping
we used $B=B_{R}=1/T$ as in \cite{LiTipErKarlAnd:JLT12}. For a fair
comparison, we also considered Nyquist-WDM systems with RZ pulses
with duty cycle 50\% opimizing also, from a spectral efficiency point
of view, the bandwidths $B$ and $B_{R}$ of the optical filters and
using $F=B$ (as mentioned, we imposed the constraint $F\ge1/T$).
The optimized values corresponding to the peak of the $\underline{\eta}_{\textrm{M}}$
curves are: $B=1/T$ and $B_{R}=0.7/T$ for 16-QAM, $B=1.1/T$ and
$B_{R}=0.85/T$ for 64-QAM, and $B=1.25/T$ and $B_{R}=0.9/T$ for
256-QAM. Filter bandwidths for each system are summarized in Table~\ref{tab:Bandwidths}.
Notice that no electric filter was considered in our simulations,
but its presence does not affect the results from a qualitative point
of view.%
\footnote{The fact that we found $B_{R}<B$ and not $B=B_{R}$ is related to
the fact that we have $F=B$ and the constraint $F\ge1/T$. Hence,
in this case it is more convenient to reduce $B_{R}$ with respect
to $B$.%
} At the receiver, since we arbitrarily chose $L_{r}=4$ for TFP-QPSK
(because corresponding to detectors with 16 states, thus with reasonable
complexity), for 16-QAM we used MAP symbol detectors taking into account
a memory $L=2$. In this way, the comparison is performed for the
same number of states of the MAP symbol detectors. On the other hand,
for 64- and 256-QAM, in order to avoid a much larger number of states,
we chose $L=1$. The Shannon limit \cite{Sh48} in the absence of
nonlinearities is also shown for comparison. It can be observed that,
in this scenario, the TFP-QPSK system outperforms the $M$-QAM systems
in spite of their higher cardinality. Although, in principle, 16-QAM,
64-QAM, and 256-QAM with polarization multiplexing could achieve spectral
efficiency values (with sinc pulses) of 8, 12, and 16 b/s/Hz, respectively,
they would be reached, in the linear regime, only for higher values
of $P/N$, whereas the optimal launch power corresponds to an SNR
value which privileges TFP-QPSK.

These information-theoretic results can be approached by using proper
coding schemes. As an example, we simulated the bit-error ratio (BER)
of a TFP-QPSK system using $B=0.325/T$, $F=0.43/T$, and employing
the rate-4/5 low-density parity-check (LDPC) code having codewords
of 64800 bits of the 2nd generation satellite digital video broadcasting
(DVB-S2) standard \cite{DVB-S2-TR}. Assuming a reference for the
BER of $10^{-7}$, the performance of this system has been reported
in Fig.~\ref{fig:SE1}. It may be observed that, despite the lack
of an optimization in the code design, we have a loss of less than
1 b/s/Hz from the theoretical results, obtaining an SE of 7.5 b/s/Hz.
The observed loss is mainly due to the presence of nonlinear effects
which require a careful redesign of the code (we used a good code
for the linear channel). We would like to mention that on this link
an experimental field trial demonstration has been also recently conducted,
reaching a spectral efficiency of more than 5 b/s/Hz despite the constraint
to use (poorly performing) 1st-order Gaussian filters, not flexible
and penalizing frequency grids, neighboring data channels with commercial
traffic, and no time available to perform the necessary system optimizations.
Similar results were obtained in a previous field trial described
in \cite{Poti:ECOC12}. The relevant results will be reported in a
later paper. 

Notice that the use of soft decoding has become a relevant topic in
coherent optical communications, and is considered to be an important
resource for performance improvement in the next future \cite{Brink:OFC12}.
Iterative detection and decoding is, at this point, a natural development
of signal processing at the receive side, as soon as the complexity
of MAP detectors (in our case only 16 states for TFP) becomes feasible
with the present technological progress, and with the intrinsic parallel
processing of block coding.

\textit{\emph{In TFP, from a conceptual point of view, the lower the
bandwidth of the optical filter and/or the frequency spacing, the
higher the peak-to-average power ratio (PAPR). However, we are optimizing
the amount of introduced packing (it is different for each value of
the launch power) so, in practice, the introduced PAPR is, in some
way, optimized. If curves in}} Fig. \ref{fig:SE1} were plotted as
a function of the transmitted power per carrier $P_{c}$ instead of
as a function of the SNR, it could be observed that TFP with QPSK
achieves the maximum value of SE, where NL effects start dominating
the performance, for a values of $P_{c}$ lower than those related
to 16-QAM systems (but higher than those related to 64- and 256-QAM
systems). This suggests that TFP with optimized packing can give values
of the optical field intensity higher than those related to 16-QAM.
So, by optimizing the amount of packing, a trade-off is reached between
the SE improvement introduced through packing and the sensitivity
to NL effects. 

Fig.~\ref{fig:SE2} reports a comparison between the same systems
as in Fig.~\ref{fig:SE1}, but compensating nonlinear fiber impairments\textbf{
}at the receive side with an ideal digital backpropagation technique
\cite{IpJLT08}, operating on the whole transmission bandwidth at
full complexity, i.e., the same complexity used to simulate the nonlinear
channel through SSFM. This will allow to increase the optimal launch
power and will allow higher order modulations to better exploit their
potential. In other words, since the channel is now ``more linear'',
we expect that Nyquist-WDM with higher order modulations outperform
TFP. This can be verified Fig.~\ref{fig:SE2} although TFP still
outperforms 16-QAM. 

We also computed the SE of the same systems but by changing the number
of sub-channels and their baud rates. In this case we compared the
different modulation formats at equal baud rate, i.e., 50 Gbaud/s,
and set the number of sub-channels in order to obtain approximately
the same total bandwidth occupation. Thus, by keeping the same filter
bandwidths as in the previous case, we set 10 sub-channels for TFP-QPSK
at 20 GHz frequency spacing, 4 sub-channels for 16-QAM and 64-QAM,
respectively at 50 and 55 GHz spacings, and 3 sub-channels for 256-QAM
at 62.5 GHz spacing. Results are very similar to those reported in
case of equal bit rate and number of sub-channels (and thus not shown
for lack of space), confirming that the TFP-QPSK performance is independent
of the granularity selected to reach the target data rate.

\begin{figure}
\begin{centering}
\includegraphics[width=1\columnwidth]{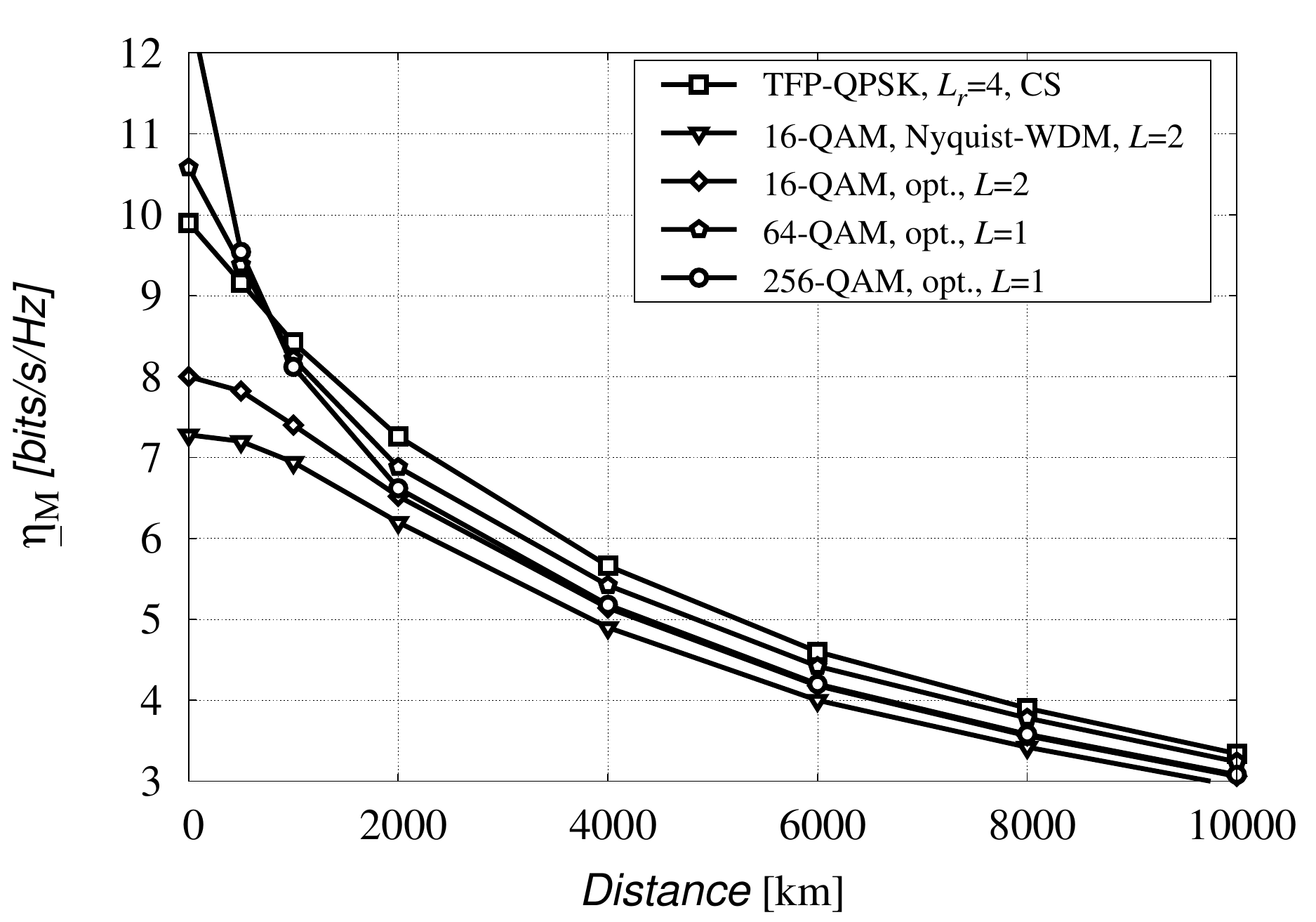}
\par\end{centering}

\caption{\selectlanguage{american}%
\label{fig:SE3}\foreignlanguage{english}{Maximum achievable spectral
efficiency of the considered systems, as a function of distance, for
all systems employing 8 140-Gbit/s subchannels.}\selectlanguage{english}%
}
\end{figure}

\begin{figure}
\begin{centering}
\includegraphics[width=1\columnwidth]{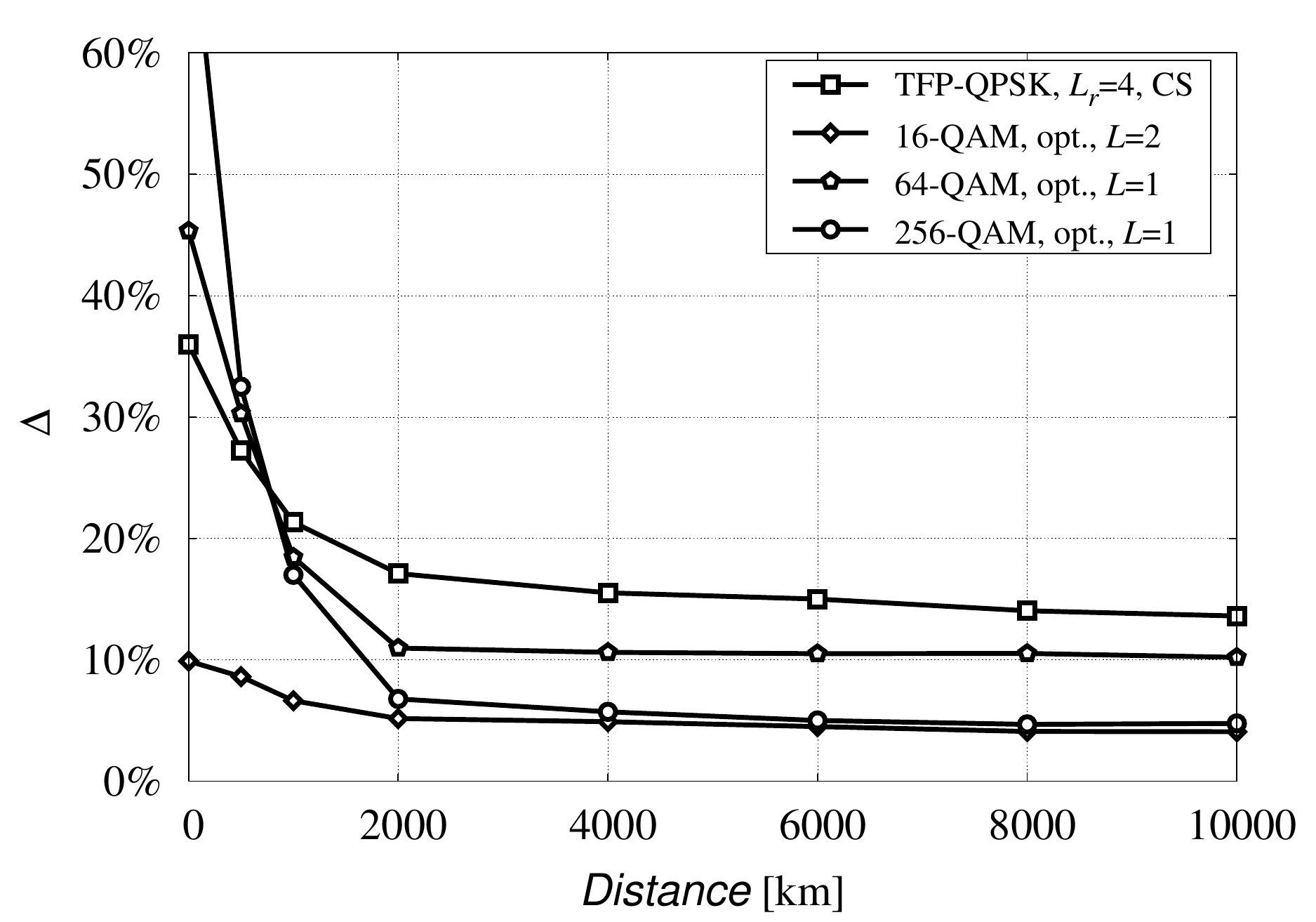}
\par\end{centering}

\caption{\selectlanguage{american}%
\label{fig:SE4}\foreignlanguage{english}{Differential normalized
spectral efficiency of the considered systems with respect to 16-QAM
Nyquist-WDM in \cite{BoCuCaPoFo11}, as a function of distance, for
all systems employing 8 140-Gbit/s subchannels.}\selectlanguage{english}%
}
\end{figure}

\begin{figure}
\begin{centering}
\includegraphics[width=1\columnwidth]{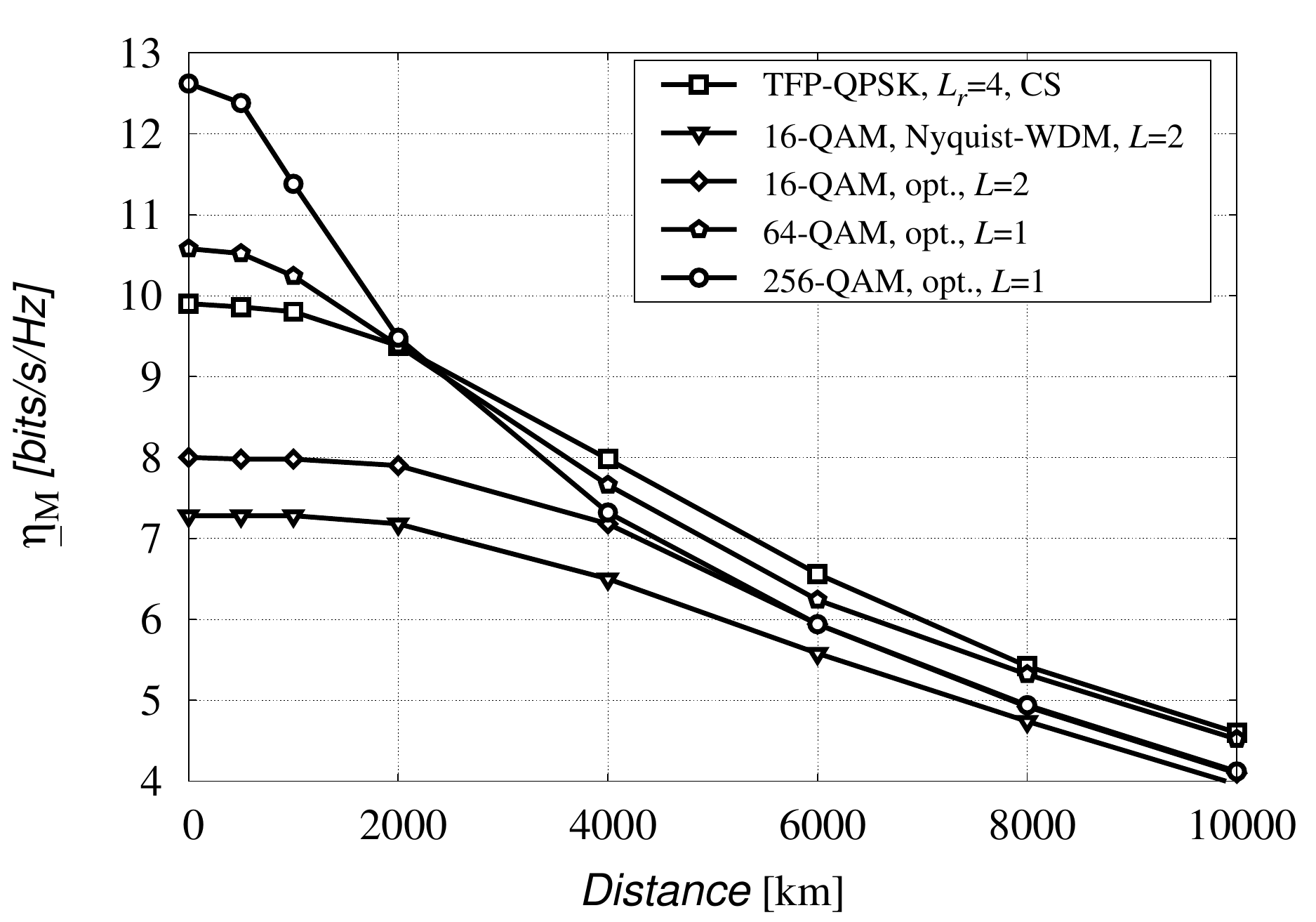}
\par\end{centering}

\caption{\selectlanguage{american}%
\label{fig:SE5}\foreignlanguage{english}{Maximum achievable spectral
efficiency of the considered systems, as a function of distance, for
all systems employing 8 140-Gbit/s subchannels, with ideal digital
backpropagation.}\selectlanguage{english}%
}
\end{figure}

In order to verify the range of applicability of the considered techniques,
in Fig.~\ref{fig:SE3} we show the maximum spectral efficiency $\underline{\eta}_{\text{M}}$
at the optimal launch power, as a function of distance on a simplified
uncompensated link (which means identical spans of length 100 km,
amplifier noise figure equal to 5 dB, no polarization mode dispersion),
for all systems considered in Fig.~\ref{fig:SE1}. TFP-QPSK is seen
to reach the highest spectral efficiency in the range 1000-10000 km,
i.e., when link nonlinear effects significantly affect signal propagation.
Fig.~\ref{fig:SE4} shows the same results but plotted as normalized
SE difference between the considered systems and Nyquist-WDM 16-QAM
in \cite{BoCuCaPoFo11}, taken as a reference, i.e., we defined
\begin{equation}
\Delta=\frac{\underline{\eta}_{\mathrm{M}}-\underline{\eta}_{\mathrm{M}}^{REF}}{\underline{\eta}_{\mathrm{M}}^{REF}}\,,\label{eq:Delta}
\end{equation}
where $\underline{\eta}_{\mathrm{M}}^{REF}$ is the maximum value
of SE achievable by Nyquist-WDM 16-QAM in \cite{BoCuCaPoFo11}. It
can be noticed that the SE gain of each modulation format becomes
constant after 2000 km, and TFP-QPSK is already the most efficient
after less than 1000 km. Fig.~\ref{fig:SE5} shows similar results
on $\underline{\eta}_{\mathbf{\mathrm{M}}}$ as a function of distance
for the systems used in Fig.~\ref{fig:SE2} (i.e., in the case of
adoption of ideal backpropagation). It can be noticed that, again,
TFP-QPSK outperforms other systems at higher distances, whereas, obviously,
in links where signal propagation occurs in the weak nonlinear regime,
the higher information rate achieved by high-order QAM systems prevails.

\foreignlanguage{american}{ }
\begin{figure}
\begin{centering}
\includegraphics[width=1\columnwidth]{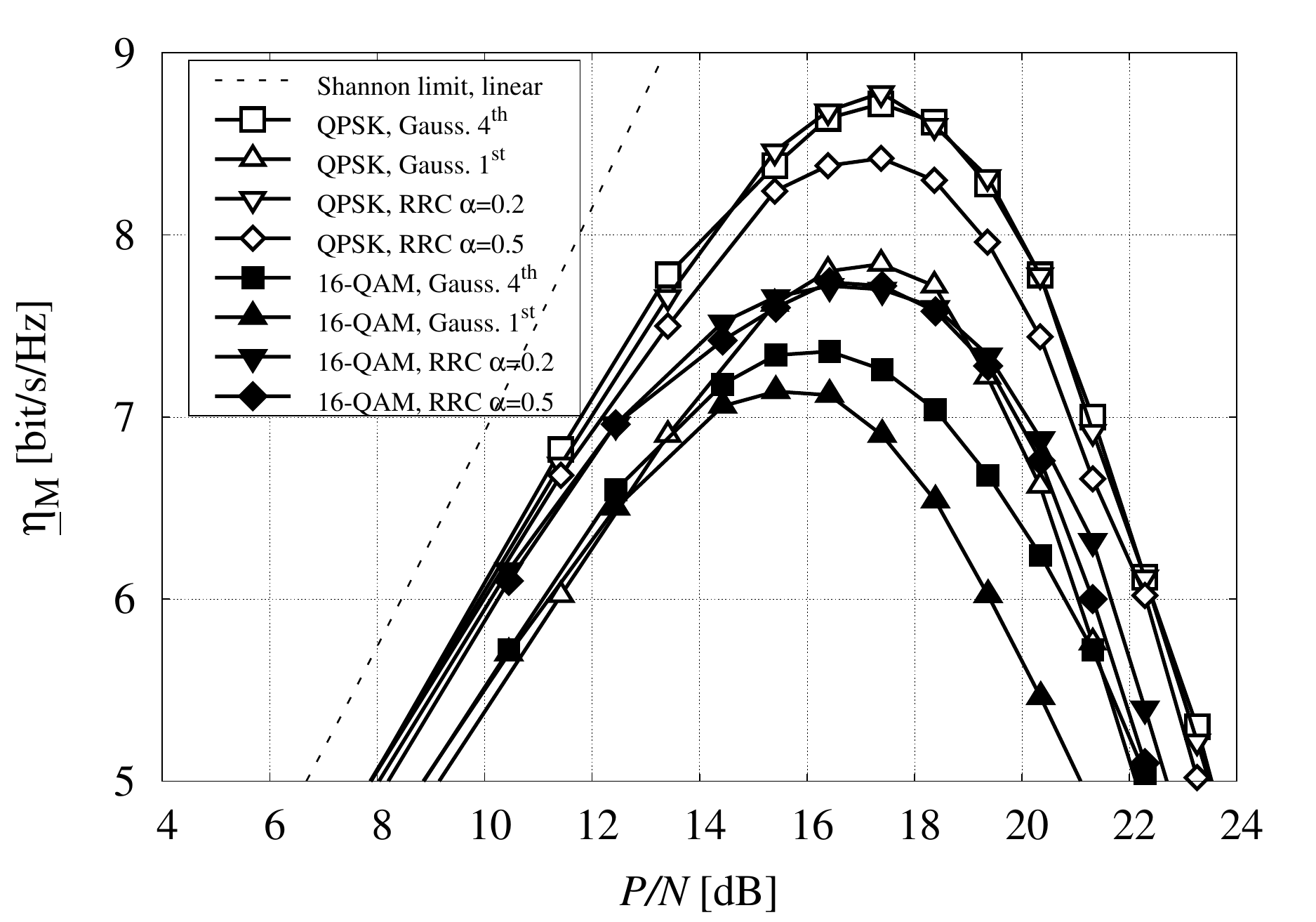}
\par\end{centering}

\selectlanguage{american}%
\caption{\label{fig:pulses}\foreignlanguage{english}{Maximum achievable spectral
efficiency as a function of different transmit pulses on a link corresponding
to 1000 km of the scenario in Figs.~\ref{fig:SE3}-\ref{fig:SE5}.}}
\selectlanguage{english}%
\end{figure}

Finally, Fig.~\ref{fig:pulses} shows the performance of TFP-QPSK
and SE-optimized 16-QAM for different transmitted pulses, on a link
corresponding to 10 spans of the scenario considered in Figs.~\ref{fig:SE3}-\ref{fig:SE5}.
In particular, in addition to the case of use of a $4^{th}$-order
Gaussian filter, we also considered the use of a $1^{st}$-order Gaussian
filter (with optimized bandwidth) and of root raised cosine (RRC)
transmit pulses (obtained through the use of proper filters) with
roll-off factor $\alpha$ equal to 0.2 and 0.5. In the case of RRC
pulses, we optimized the frequency spacing $F$ and the symbol time
$T$ (related to the frequency support $B$ of each subcarrier by
$B=(1+\alpha)/T$). Results suggest that the pulse deriving from the
adoption of a $4^{th}$-order Gaussian filter performs almost as good
as the RRC pulse with roll-off 0.2, whereas a significant penalty
is observed when using a $1^{st}$-order Gaussian filter. Nevertheless,
in all cases TFP-QPSK performs better than 16-QAM under this scenario.

\section{Conclusions\label{sec:Conclusion}}

We compared different techniques to improve the spectral efficiency
of long-haul optical systems. With the exception of short links where
signal propagation occurs in the weak nonlinear regime, the most promising
solution is shown to be that based on time-frequency packing which
is related to the use of narrow optical filtering and a tight packing
of the carriers in frequency, giving up the signal orthogonality in
the time and frequency domains, and on the adoption of detectors able
to cope with the interference intentionally introduced in the system.
When nonlinearities start dominating the performance, this solution
provides better results than other solutions proposed in the literature
and based on higher-order modulations, showing that, when nonlinear
effects are present, the spectral efficiency cannot be trivially increased
by increasing the modulation order. This result is confirmed also
for ultra-long-haul links, up to 10000 km, whereas better SE values
can be achieved using modulations with very high cardinality, such
as 256-QAM, but only for short range links where fiber nonlinearities
have a weak effect, or resorting to compensation techniques such as
digital backpropagation. We also reported simulation results on a
modulation and coding format which, on a realistic optical link, reaches
a spectral efficiency of 7.5 b/s/Hz with a polarization-multiplexed
time-frequency-packed QPSK, with a loss of less than 1 b/s/Hz from
the information-theoretic results.

\section*{Acknowledgment}

This work was supported in part by CNIT and by the Italian Ministero
dell'Istruzione, dell'Universit\`a e della Ricerca (MIUR) under the
FIRB project Coherent Terabit Optical Networks (COTONE). The authors
would like to thank the Editor, Prof. Erik Agrell, and the Reviewers;
their comments and constructive criticism helped to improve the quality
of the paper.


\begin{IEEEbiography}[{\includegraphics[width=1in,height=1.25in,clip]{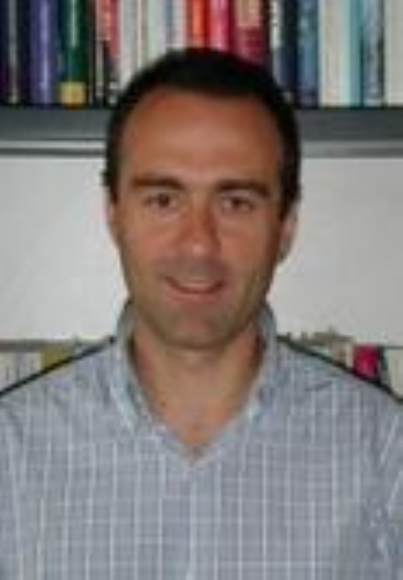}}]
{Giulio Colavolpe} (S'96-M'00-SM'11) was born in Cosenza, Italy, in 1969. He received the Dr. Ing. degree in Telecommunications Engineering (cum laude) from the University of Pisa, in 1994 and the Ph.D. degree in Information Technologies from the University of Parma, Italy, in 1998. Since 1997, he has been at the University of Parma, Italy, where he is now an Associate Professor of Telecommunications. In 2000, he was Visiting Scientist at the Institut Eur\'ecom, Valbonne, France. His research interests include the design of digital communication systems, adaptive signal
processing (with particular emphasis on iterative detection techniques for channels with memory), and information theory.

He received the best paper award at the 13th International Conference on Software, Telecommunications and Computer Networks (SoftCOM'05), Split, Croatia, September 2005, the best paper award for Optical Networks and Systems at the IEEE International Conference on Communcations (ICC 2008), Beijing, China, May 2008, and the best paper award at the 5th Advanced Satellite Mobile Systems Conference and 11th International
Workshop on Signal Processing for Space Communications (ASMS\&SPSC 2010), Cagliari, Italy.  He is currently serving as an Editor for \textit{IEEE Transactions on Communications} and \textit{IEEE Wireless Communications Letters}. He also served as an Editor for \textit{IEEE Transactions on Wireless Communications} and as an  Executive Editor for \textit{Transactions on Emerging Telecommunications Technologies (ETT)}.
\end{IEEEbiography}

\begin{IEEEbiography}[{\includegraphics[width=1in,height=1.25in,clip]{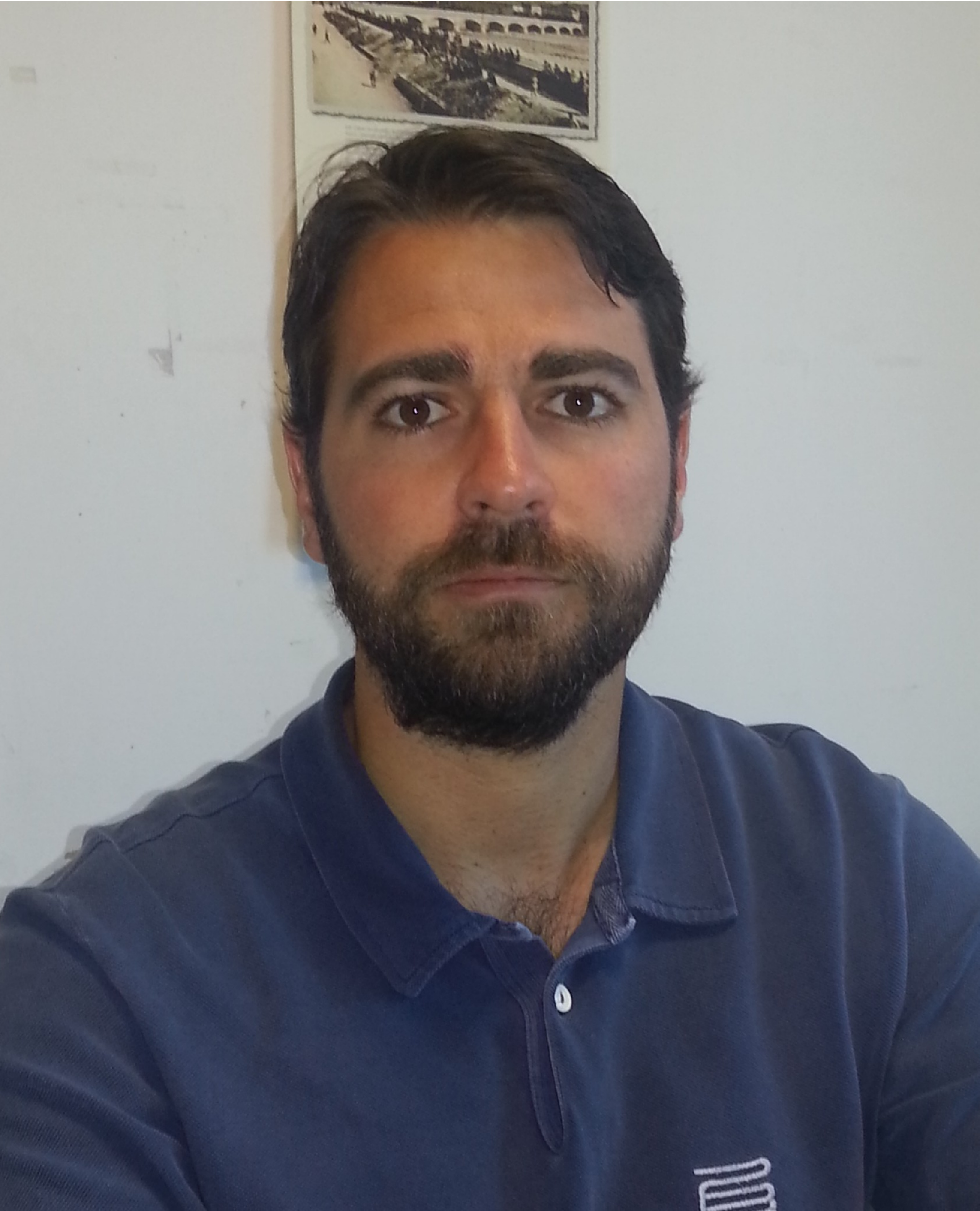}}]
{Tommaso Foggi} was born in Parma, Italy, in 1978. He received the master's degree in Telecommunication Engineering from the University of Parma in 2003 and the Ph.D. degree in Information Technology from the same University in 2008.
Since 2009 he is a research engineer of National Inter-University Consortium for Telecommunications (CNIT) in the University of Parma research unit.

His main research interests include electronic signal processing for optical communication systems, in particular adaptive equalization, coding and iterative decoding techniques, optical channel impairment compensation, channel estimation, 
simulative software implementation.
He is author of several internation journal and conference papers and patents, and he is recipient of the the best paper award for the Optical Networks and Systems symposium at the IEEE International Conference on Communcations (ICC 2008), Beijing, China, May 2008.
He has been involved in research projects funded by the Italian Ministry of University and Research (MIUR), by the European Space Agency (ESA), 
and by private companies (Marconi-Ericsson).
\end{IEEEbiography}

\end{document}